Universal Law for the Elastic Moduli of Solids and Structures

by

S. J. Burns[1] and Sean P. Burns[2]

[1] Materials Science Program

Department of Mechanical Engineering

University of Rochester

Rochester, NY 14627-0132, USA

[2] Mesoscale and Microscale Meteorology Division

National Center for Atmospheric Research

3090 Center Green Drive

Boulder, Colorado 80301, USA

June 7, 2020

*Abstract*

A law previously found for shear moduli of crystalline materials is developed and extended to all elastic moduli in solids and structures. Shear moduli were previously shown to depend only on specific volume. The bulk moduli of many materials and structures are now predicted analytically and empirically shown with unerring accuracy by observing the elasticity as a specific volume power law. The law is supported by experimental evidence from: foams, Schneebeli 2-dimensional graphene mats, metamaterials, fully dense metals, ceramics and minerals. This new, generalized, universal, elastic moduli law always describes materials that support shear stresses i.e., solids; it is shown that all elastic moduli are directly dependent only on the specific volume.

*Introduction*

The elastic properties of solids have been of interest for a very long time. Over 70 years ago there was a series of papers that investigated the hypothesis that the elastic moduli of single and polycrystalline solids are only dependent on the specific volume [1, 2]. Both papers concluded that the experimental evidence at that time did not support the hypothesis. More recently empirical evidence using a semi-log plot from a wide selection of materials establishes that the moduli are only dependent on changes in volume due to temperature and pressure [3]. Change in volume plus a reference volume gives the specific volume of the material in reference [3]. The implication is that this work supports the hypothesis that the elastic bulk moduli depend only on the volume up to specific volume changes of 30 to 40% in volumetric strains.

A new power-law relation between the elastic constants and the specific volume seems to have several unconnected origins with a very wide range of applicability: Gibson and Ashby working with a foam called the relation between modulus and density a 'scaling-law' with the power exponent between 2 and 3 for open and closed cell foams respectively [4-7]; O. L. Anderson's relation with thermal expansion which is dependent on the 'Anderson-Grüneisen' exponent with the power-law exponent typically about 5 for fully dense solids [8-11]; Burns related the elastic shear compliance to the specific volume using thermodynamics and called his relation a 'constitutive-law' [12, 13]; Birch and then Murnaghan noted that an 'Equation of State' found by Birch-Murnaghan can be reduced to a simple power-law relating the modulus to the density through a power-law in most cases [14, 15]. Grover, Getting and Kennedy as noted above called their semi-logarithm form, which is an engineering approximation to a power law, a 'Simple Compressibility Relation for Solids' and applied it to over 70 solids [3]. Finally, the elastic relation for graphene, man-made metamaterials at the limits of Hashin-Shtrikman structures and 2-dimensional Schneebeli materials are again described by a power-law [16-21]. It seems this law is 'universal' which is expanded upon and supported below.

There was a serious oversight made in thermo-elasticity from about a century and half ago that has direct applicability to the power-law relation noted above: Neumann's Principle relates *the physical properties to the symmetry elements of the point group of the crystal class* thus the single crystal's symmetry dictates the polycrystalline, thermal, mechanical, electrical, optical and magnetic physical properties [22-25]. In stating his principle, Neumann failed to recognize that shear thermal expansion coefficients were in general not always zero. Most crystal classes except for the lowest symmetry were considered to always have zero thermal expansion coefficients in shear. This error was repeated by his student, Voigt, and then by colleagues in the same department including Born, Debye, and later in the literature by Brillion, Landau, Nye, Newnham, Wallace, deWit, Kim, Barron, Burns and essentially all others who reinforced this incorrect concept [13, 22-28]. With time, it has become fully embedded in the high-symmetry single crystal and polycrystalline thermodynamic literature as a fact: a researcher today would believe that the shear thermal expansion coefficient is zero as most materials will not change shape upon heating. Yet, without a shear thermal expansion coefficient the shear phonons can't couple into the shear part of the entropy. These same researchers noted above would support an engineering shear strain, $\gamma$ due to a shear stress, $\tau$ as seen in equation (1). $\mu_T$ is the isothermal shear modulus and $\lambda_T$ is the isothermal shear compliance. $T$ is the absolute temperature.

$$\gamma = \frac{\tau}{\mu_T} = \lambda_T \tau \; ; \; \lambda_T = \frac{1}{\mu_T} \tag{1}$$

The temperature derivative of the shear strain in equation (1) at constant shear stress is the shear thermal expansion coefficient displayed in equation (2) and graphically shown in figure 1 (a):

$$\left.\frac{\partial \gamma}{\partial T}\right|_\sigma = \tau \left.\frac{\partial (1/\mu_T)}{\partial T}\right|_\sigma = \tau \left.\frac{\partial \lambda_T}{\partial T}\right|_\sigma. \tag{2}$$

$\sigma$ represents all other stresses. Thermal expansion coefficients here are defined as the thermal derivatives of strain at constant stress, not necessarily at zero stress, as has been assumed in the literature. Solids with an applied shear stress will always change shape upon heating as seen in figure 1 (b) and explained below.

Recent work using third-order derivatives of the Gibbs function investigated such a relationship and found that some thermodynamic identities are inconsistent with experimental data since the shear modulus and density both decrease with $T$ so the product can't in general be constant [13]. It was recognized then that the assumption of zero shear thermal expansion coefficient was not correct except at zero shear stress. The physics of shear thermal expansion coefficients is clear: an isotropic solid will change shape in response to an applied shear stress through the $T$ dependence of the shear compliance as seen in equation (2).

In all crystalline solids, including single crystals, the thermal expansion coefficients, $\alpha_{ij}$, follows the thermal part of the strain tensor, $\varepsilon_{ij}$. The components of the strain tensor are proportional to a small change in temperature $\Delta T$ so the tensorial thermal expansion can always be diagonalized as shown below.

$$\varepsilon_{ij} = \alpha_{ij} \Delta T \quad \text{so} \Rightarrow \begin{pmatrix} \alpha_{11} & 0 & 0 \\ 0 & \alpha_{22} & 0 \\ 0 & 0 & \alpha_{33} \end{pmatrix} \tag{3}$$

A general symmetric strain tensor has three principal, normal, eigenvalue strains and therefore the most general thermal expansion coefficients in crystalline solids are the three independent, thermal expansion coefficients, $\alpha_{ii}$, given in equation (3). The shear strain thermal expansion coefficients are apparently zero as seen in expression (3).

The tensorial shear properties are found from the differences in the principal Cauchy values and are the deviatoric thermal expansion coefficients in isotropic materials in a rotated coordinate system. In all crystal classes, the thermal expansion coefficients conform to Neumann's Principle and when the coordinates are the principal orthogonal system not necessarily aligned with the crystal's principal axes, the shear thermal expansion coefficients in the aligned system are necessarily zero [22-25] with zero applied shear stress. Figure 1 (b) shows material schematically in the first panel labeled (i) on left it is at a temperature $T_0$. The square is then heated with the temperature change of $\Delta T$. The material will expand following expression (3) on the right side and is now displayed as a rectangle on the second upper panel (ii) of the schematic. The material does not change shape if it is heated using the principal coordinates of figure (3). Figure 1 (b) shows that in the lower left panel (iii) the application of a shear stress, one of the of the three independent shear stresses, causes the material to change shape. The square is now a rhombus. Panel (iii) shows the shape change due to the mechanical application of a shearing stress. Upon heating by $\Delta T$ the shape again changes as seen in the lower right panel (iv) which shows a larger shear strain again in a rhombus; the same result is seen in figure 1 (a) as the shear thermal expansion coefficient follows the green arrow; this is also seen in equation (2). The application of shear stresses is the source of the shear thermal expansion coefficients which are only zero when the shear stresses are zero as seen in equation (2) and panel (ii).

No distinction is usually made between the isothermal and isentropic shear moduli, $\mu_T$ and $\mu_s$. $s$ is the entropy used in the shear modulus subscript. Shear, stress-wave speeds for over a century weren't corrected for thermal effects because the shear thermal expansion was always considered to be zero. The stress-wave speeds are so fast there would not be adequate time to allow heat to equilibrate over the stress' wave length. The presumption of no shear thermal expansion negates the necessity to consider heat; there is no temperature change, so there is no heat to equilibrate.

*Investigation of Adiabatic and Isotherms Shear Based Systems*

In the shear stress-strain system, the shear thermal expansion coefficient follows equation (2) and is zero only at zero shear stress; the adiabatic slopes were always presumed to align with the isotherm slopes as is noted above and seen in figure 2. Figure 2 shows two assumed linear adiabatic shear lines passing through the point $\tau = 0$; $\gamma = 0$. This results from the shear thermal expansion coefficients having always been thought to be zero. The zero-shear thermal expansion mentioned above led to the incorrect conclusion that with shear elastic moduli, the isotherms and adiabatics are co-aligned where the isotherms cross with $\mu_T = \mu_s$. The adiabatic lines can't follow the isotherms through the point of zero shear stress and strain because the isentropic lines will then touch and from Carathéodory's principle there will be a violation of the second-law of thermodynamics. The second-law violation is expanded on below and shown using figure 2. Figure 2 demonstrates that $\mu_T = \mu_s$ is not possible since it leads to a construction of two hypothetical adiabatic lines connected by a single isotherm. The cycle is necessarily three-sided with the isentropic lines following the isotherms through the point $\tau = 0$, $\gamma = 0$. An isotherm from $1 \rightarrow 2$ connects the isentropic lines and thus there is mechanical work in the internal energy-like cycle shown in the figure. Figure 2 and equation (4) establish a Kelvin- Planck second law contradiction. The cycle is:

$$-\oint \tau d\gamma = \oint T ds = T(s_2 - s_1) = T\Delta s \qquad (4)$$

The Kelvin-Planck statement of the second-law is violated since the cycle takes heat from a single source and generates negative mechanical work on the left side of equation (4) and is seen in figure 2. The mechanical work is shown in red. If two isentropes were to touch then a three-sided cycle can always be constructed in violation of Carathéodory's principle. The cycle in figure 2 can't exist because the isentropic lines cross at $\tau = 0$, $\gamma = 0$. Figure 2 shows that such a supposition is in direct conflict with the Kelvin-Plank statement of the second law and Carathéodory's principle and is thus not possible.

Figure 3 is the published data for the $T$ dependence of the shear compliance $\lambda_T$ of polycrystalline copper. The measured data [29, 30, 12] was $\mu_T$ and is used below to construct an adiabatic line found by an energy balance in a two-independent-variable thermodynamic system; it assumes that a simple power series in $T$ describes the temperature dependence of $\lambda_T$.

*Energy Balances in Sheared Elastic Solids*

The incremental energy balance for the thermodynamic system with an independent shear stress, $\tau$ and engineering shear strain, $\gamma$, is given in equation (5).

$$du = Tds + \tau d\gamma \qquad (5)$$

$u$ is the internal energy in the system. The Gibbs-like free energy at constant shear stress is $g$.

$$g \equiv u - Ts - \tau\gamma . \qquad (6)$$

So from (5) and (6),

$$dg = -sdT - \gamma d\tau . \qquad (7)$$

We see from equation (7) that:

$$\left.\frac{\partial g}{\partial T}\right|_\tau = -s \; ; \; \left.\frac{\partial g}{\partial \tau}\right|_T = -\gamma \qquad (8)$$

and from equation (8), we obtain the Maxwell relation:

$$\frac{\partial^2 g}{\partial \tau \partial T} = -\left.\frac{\partial s}{\partial \tau}\right|_T = -\left.\frac{\partial \gamma}{\partial T}\right|_\tau = \frac{\partial^2 g}{\partial T \partial \tau} \qquad (9)$$

The middle expression in (9) gives a differential expression for the entropy due to the shear stresses.

$$ds = \left.\frac{\partial \gamma}{\partial T}\right|_\tau d\tau \qquad (10)$$

Substitution from equation (2) which assumes a linear system, into equation (10) and integration with $T$ constant as dictated by the middle-left side of equation (9) allows for integration and yields $s$ for linear elastic systems. The derivative of the shear compliance is a direct result of equation (1) being linear.

$$s = \frac{\tau^2}{2} \left.\frac{\partial \lambda_T}{\partial T}\right|_\sigma \qquad (11)$$

A constant $s$ curve in $\tau$ versus $\gamma$ space is obtained by knowledge of the $T$ dependence of $\lambda_T$. This is easily seen by solving equation (11) for $\tau$ with $s$ constant; it yields:

$$\tau = \left( \frac{2s}{\left.\frac{\partial \lambda_T}{\partial T}\right|_\sigma} \right)^{1/2} \qquad (12)$$

So with the aid of equation (1) we find:

$$\gamma = \lambda_T \left( \frac{2s}{\left.\frac{\partial \lambda_T}{\partial T}\right|_\sigma} \right)^{1/2} \qquad (13)$$

The $T$ dependence of $\lambda_T$, $\left.\frac{\partial \lambda_T}{\partial T}\right|_\tau$ and a fixed value for $s$, yields a parameterized curve for $\tau$ versus $\gamma$ with entropy constant.

The shear modulus versus $T$ has been measure extensively in the literature. Data for copper covers a very large temperature range. Figure 3 shows $\lambda_T = \frac{1}{\mu_T}$ versus $T$ patched from two measurements in [29, 30, 12]. The data was recorded as $\mu_T$ versus $T$. These two different measurements were matched at $T = 300\text{K}$. The higher $T$ values have been corrected with the ratio of the match at $T = 300K$ to result in a smooth curve. Figure 3 shows a third-order power series fit to the measured data.

Figure 4 is a construction of the constant entropy curve using equations (12) and (13). The third-order power series fit to the data in figure 3 was used to construct the $T$ dependence of $\lambda_T$ and $\left.\frac{\partial \lambda_T}{\partial T}\right|_\sigma$ in the parametric expressions for the constant entropy curve. The isentropic slope of figure 4 versus $T$ is plotted in figure 5. This slope is the isentropic shear modulus, $\mu_s$ i.e., the slope following the constant $s$ curve. The isothermal shear modulus, $\mu_T$, taken from the original data is also given for comparison in figure 5.

There is thermodynamic relation that relates the isothermal and isentropic shear moduli. The relation can be found using Jacobian algebra, Table 1, and is given in expression (T-2).

$$\lambda_s = \lambda_T - T\left(\frac{(\tau \frac{\partial \lambda_T}{\partial T}|_\sigma)^2}{C_\tau}\right) \qquad (14)$$

The constant shear stress heat capacity is $C_\tau$ for this shear stress system. It is included in Table 1. $C_\tau$ is found with the aid of equation (11) in Table 1. Using the expression for the heat capacity in Table 1, in equation (14), the following expression is just equation (14) with the primes as the first and second derivatives:

$$\lambda_s = \lambda_T \left(1 - 2\frac{(\lambda_T')^2}{\lambda_T \lambda_T''}\right) \qquad (15)$$

*A Universal Law for the Volume Dependence of the Shear Modulus*

There is one form for the shear modulus that will allow isothermal lines to exist at the zero stress crossing. The requirements are for all isothermal crossings to have no second-law violations, no temperature changes in the sheared solid except through $v$, and keep $\mu_s$ proportional to $\mu_T$ so a solid can be rapidly sheared without changing temperature on crossing through the zero-thermal expansion coefficient point. This form is possible if and only if, the differential equation in the right side of equation (15) is satisfied for the adiabatic lines at the isothermal crossing point and throughout the shear-stress space [12]. The $T$ derivatives of the shear compliance will not allow these conditions to occur as seen in figures 2 and 5. The differential equation below and on the right of equation (15) describes the slope of adiabatic stress versus strain curve at all points in the entire $\tau$ vs $\gamma$ space. Take the shear compliance to be $\lambda(v)$ and primes on $\lambda$ are derivatives with respect to *v* so the adiabatic slope is proportional to the isothermal slope. The shear compliance is not the same as $\lambda_T$ where the elastic compliances are in general considered to be functions of the thermodynamic variables. $\lambda$ is considered only as a function of *v* which avoids the difficulties seen in figures 2 and 5. The numerical values of $\lambda$ and $\lambda_T$ are identical with $\lambda = \lambda(v)$.

$$\frac{(\lambda')^2}{\lambda \lambda''} = \text{constant}. \tag{16}$$

The general solution to the differential equation (16) is from Appendix A, equation A-12:

$$\frac{\lambda}{\lambda_0} = \left(\frac{v}{v_0}\right)^m; \quad \lambda = C_0 v^m \text{ or } \mu = \frac{\rho^m}{C_0}. \tag{17}$$

In equation (17) *v* is the specific volume, $v_0$ is a reference volume at the same reference state as the shear compliance, $\lambda_0$; $\rho$ is the density and $\mu$ which only depends on *v* is Lame's second shear modulus and is now only a function that depends on *v* i.e., a constant volume line. *v* in turn is a function of the pressure *p* and *T* and all other thermodynamic variables. Equation (17) when substituted into (16) yields a constant on the right side of equation (16) with m as a constant. m is not one of the two arbitrary constants from integration of the second-order differential equation (16); the two constants of integration are given as seen in Appendix A.

*All elastic constants must depend only on the specific volume*

The argument to mandate all elastic constants only depend on specific volume is based on the transformation laws of elastic constants in one orientation to elastic constants in a second rotated coordinate system i.e., the *state of stress* in a tensor stress field. Rotations depend on the direction cosines between the two coordinates systems and the other elastic constants. In isotropic materials a $\pi/4$ coordinate rotation is well known to relate the shear modulus in the rotated coordinate system to Young's modulus, *E*, and Poisson's ratio, $\upsilon$, in a uniaxial stress state. *E* is obtained from the uniaxial stress versus uniaxial strain and Poisson's ratio which is the uniaxial strain proportioned to the negative strain on a stress free perpendicular coordinate.

$$\mu_T = \frac{E}{2(1+\upsilon)} \tag{18}$$

Generally, *E* and $\upsilon$ are considered as functions of *T*, *p* and other thermodynamic variables. However, equation (17) has established that the shear modulus is only a function of *v*.

$$\mu(v); v = v(T, p) \tag{19}$$

Expression (18) will only be valid if $\mu_T = \mu$. Thus, the left side of (18) is only a function of a single variable, $v$, and the right side is considered a function of all the thermodynamic variables including $T$ and $p$ in both $E$ and $\upsilon$. The *state of stress* requires that both sides of (18) be functions of the same variable or variables and not be dependent on the choice of the coordinate system. The conclusion for the moduli transformation in expression (18) to be valid is

$$E = E(v) \text{ and } \upsilon = \upsilon(v) \qquad (20)$$

In a crystal, the tensor transformation of the elastic moduli, $c_{noqr}$, follows from expressions from reference [23]

$$\mu = c_{ijkl} = a_{in} a_{jo} a_{kq} a_{lr} c_{noqr} \qquad (21)$$

The shear modulus, $\mu$ on the left hand side of equation (21) is only a function of $v$. The other 21 elastic constants in the general case are thought to be $T$ and $p$ dependent; the $a_{ij}$ are the direction cosines on the right hand side of (21). Consider the left side of (21) to be a shear modulus which only depends on $v$. Then $c_{noqr}$ must also depend only on volume since the direction cosines are numbers. Equation (21) is the tensor transformation between rotated coordinates; in tensor notation $i$, $j$, $k$ and $l$ go from 1 to 3 as does $n$, $o$, $q$ and $r$. The tensor notation uses sums of the components in equation (21). Again, as shown in equations (18) through (20), all the elastic moduli must only be dependent on the specific volume, $v$ so $c_{noqr} = c_{noqr}(v)$.

It follows that all crystalline elastic stiffnesses and the elastic compliances are therefore not thermodynamic functions; like the shear modulus they must be functions only of the specific volume. They can't be thermodynamic expressions dependent on additional variables. Solids support stresses but fluids don't; stresses are tensors so equations (18) through (21) follow for solids and describe the tensor transformations of both elastic stiffnesses and compliances. Elastic moduli in crystals or effective moduli in solids and complex microstructures must only be functions of the specific volume. Empirical support of this statement is given below while references [3-21] have already noted moduli depending only on $v$.

*Empirical Evidence to Support the Specific Volume Dependence of the Elastic Moduli*

Empirical evidence will now be presented in support of equation (17). The first materials selected are from shear moduli of metals, ceramics and minerals. Figure 6 throughout uses relative values for both the elastic shear compliances and volumes. The investigation on copper used the $T$ dependence of $\lambda_T$ and $v$ in equation (17) in parametric form. Figure 6 is a plot of $ln(\lambda_T)$ *versus* $ln(v)$ for oxygen-free copper. The assumption in the data is that $\mu_s = \mu_T = \mu$. The data were taken from reference [29]. Each point in this plot is at a different $T$ or $p$. The data are from very low temperatures near 5K to about 300K. Thus, all quantum effects at low temperatures are included in the data presented. *v* in the data used in figure 6 is from numerical integration of 3 times the measured linear thermal expansion coefficient. The material used in the volume data and in the shear modulus is identical i.e., from the same copper sample.

MgO single crystal is the second material in figure 6 that was chosen for study [31]. The data are again for thermal changes but now for a single crystal's elasticity. The single crystal shear compliance $S_{44}$ scale is also on the left of the graph. The temperature goes from $T$=300K to $T$=1,800K. The shear-compliance $S_{44}$ versus volume with pressure as a parameter is also shown in figure 6. The elastic properties in compression were measured by Brillion spectroscopy in a sample within a diamond anvil cell. The material in the volume data and in the shear modulus data is from the same single crystal [32]. Figure 6 shows a plot of $ln(S_{44})$ *versus* $ln(v)$ using parametric *p* and *T* values in the elastic shear compliance, $S_{44}$ and *v*. The density comes from X-ray measurements of interatomic plane spacings. Figure 6 on the left scale shows a plot of $ln(S_{44})$ *versus* $ln(v)$ using relative values but it at a slightly different reference state from the thermal data. This pressure data is in excellent agreement with the thermal data as noted above. The pressure data have significantly more scatter than the more precise thermal data. Also, increasing the temperature increases *v* while increasing the pressure decreases *v*. This is seen by having the temperature data in the 1$^{st}$ quadrant and the pressure data in the 3$^{rd}$ quadrant in the plot as is expected physically for *v*.

Olivine is the third material chosen for shear modulus investigation. The plot in figure 6 is of $ln(S)$ *versus* $ln(v)$ again with *T* as a parametric variable. The data for this mineral were taken from reference [33]. The temperature goes from 300K to 1,400K. The material used in the volume data and in the shear compliance data are again the same material. Olivine is a two phase material within a solid-solution phase field.

Figure 7(a) and 7(b) are plots of the elastic compressibility or the reciprocal bulk modulus of materials versus specific volume. Again, the linear fit to the measured data all have R values of 0.997 or better. The implication is that the dilatational components of the Cauchy stress tensor which are described by equations (18-20) empirically follow equation (17). Figure 7 also shows several other minerals and metals.

The restriction on equation (17) is that the solid must support a shear stress. The solution to expression (16) should also apply to all material selections with very different nano, micro or macro structures. The recent design of material structures designed to be very stiff yet extremely light weight are reported [17, 18]. Metamaterials, engineered foams structures with both open and closed structures, graphene mats with fiber structures, etc. all seem to be described by equation (17). Figure 8 is a plot of the log natural of 1/(effective shear modulus) versus the log natural of the volume from reference [17]. The graph is the meta-structures of: Cubic foams, Octet foams, Cubic + Octet, Cubic + Octet 2, Octet Truss, Isotropic Truss and Quasi-random foams. The meta-material foam curves have been fit as straight line data in log-log plots with m between 1.07 and 1.28 respectively for the measured power exponents.

*Conclusions*

The universal equation proposed in this manuscript starts with classical thermodynamics applied to sheared elastic solids. In particular, considerations of the elastic shear moduli at all temperatures and pressures were shown to be through the material's density or volume. The bulk elastic moduli investigated here also only depend on $v$ through a power law. The universal equation proposed theoretically relies on the isothermal shear stresses and shear strains passing through a point that includes all isotherms and pressures. This is also the point of zero shear thermal expansion coefficients at all temperatures and pressures in single crystals or polycrystalline materials. The material does not have to be crystalline but it must support shear stresses i.e., it is restricted to a solid. The adiabatic shear-stress versus shear-strain curves are repelled by the point of zero shear thermal expansion coefficients and can't follow the isotherms through this point because of second law of thermodynamic considerations. It is argued here that there is only a single form for all elastic moduli that avoids second-law violations: it relates the material's specific volume $v(T, p)$ to the modulus for isotropic materials for shear, $\mu(v)$. $\mu(v)$ has a thermodynamic dependence on $T$ and $p$ only through $v$. The $v$ dependence of the bulk moduli presented here is through empirical evidence and supporting equations (18-21). The same behavior is observed for effective bulk moduli in solids and structures. This change in elastic properties has significant implications for many elastic-based mechanical properties and our understanding of elastic solids.

*Acknowledgements:*

SJB would like to thank: J. C. Lambropoulos, Ranga Dias, R. M. McMeeking, J. B. Berger, D. N. Polsin, G. W. Collins, and J. R. Rygg for discussions and the Department of Energy through Basic Energy Sciences, Materials Science and Engineering who supported earlier work related to this topic. SPB would like to thank the National Center for Atmospheric Research (NCAR) which is funded through NSF.

*Figures*

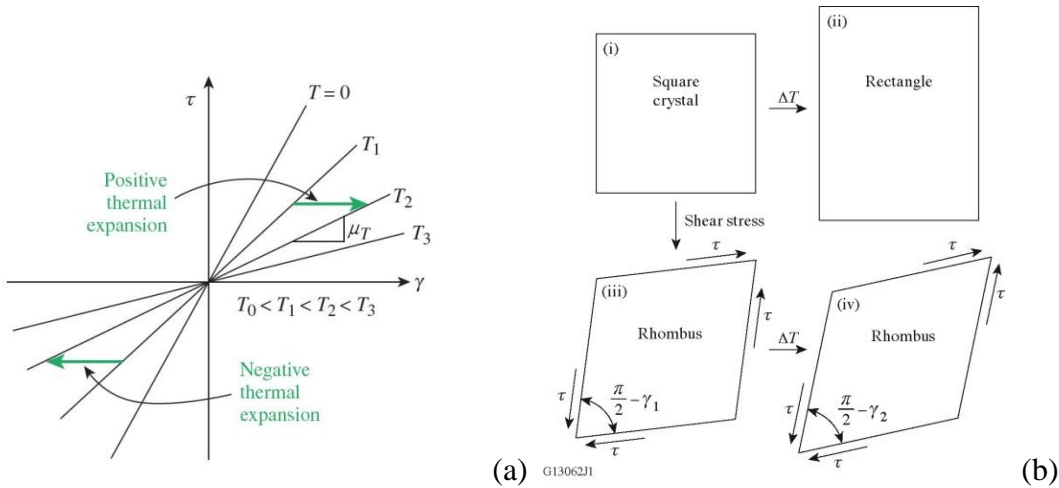

Figure 1(a). A schematic showing isothermal, linear-elastic, shear-stress, $\tau$, versus shear-strain, $\gamma$, lines. All the isotherms cross at the point (0, 0). The shear thermal expansion is marked in green. (b) A schematic of thermal principal strains seen in material with and without applied shear stress. The first panel (i) is a square; the second panel marked (ii) is a rectangle after heating, see equation (3); the third panel marked (iii) is a rhombus with the application of shear stress; the last panel marked (iv) is a rhombus with a more acute shear-strain angle when it is sheared and heated.

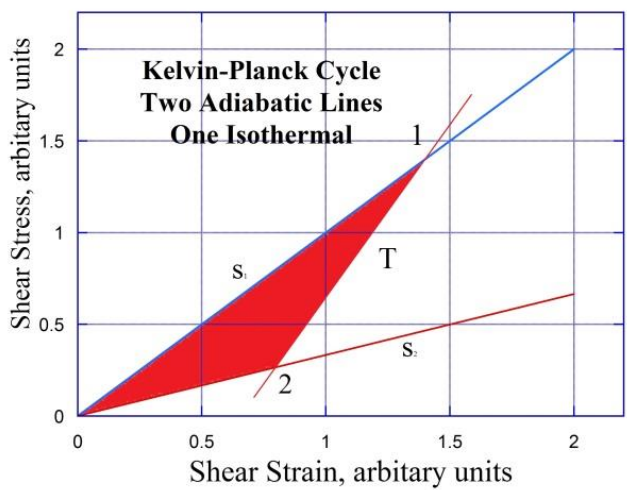

Figure 2 A thermodynamic cycle in contradiction to the second-law of thermodynamics is shown. The isotherm intersects the adiabatic line $s_1$ at point 1 and adiabatic line $s_2$ at 2. The complete three-sided cycle $0 \to 1 \to 2 \to 0$ including one isotherm and two adiabatic lines is shown. The isothermal slope is steeper than the adiabatic slope as shown.

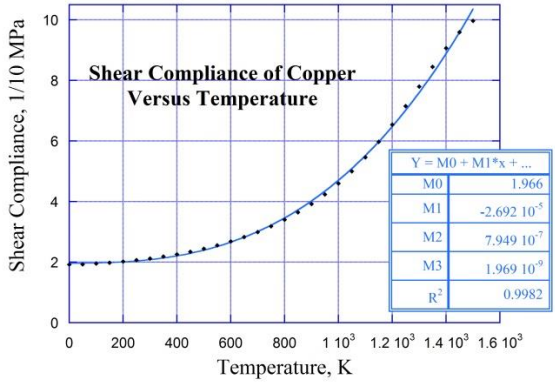

Figure 3. Shear compliance of copper versus *T*. A third order curve fit to the data is shown. The original data is from references [29, 30, 12]. The copper melts at 1357 K. See [12] for details.

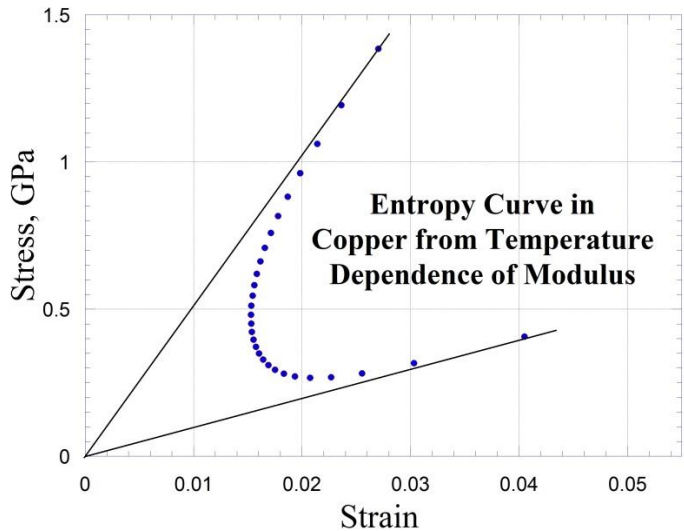

Figure 4. The shear stress versus shear strain on a constant entropy curve with two isotherms is displayed. The mirror image is in the 3$^{rd}$ quadrant. *s* was chosen as 3.3 J/mole*K, the density of copper was taken as 8.96 gm/cm$^3$ and the atomic weight as 63.55 au. Equations (12) and (13) plus the curve fit data from figure 3 are used for the construction of the dotted curve.

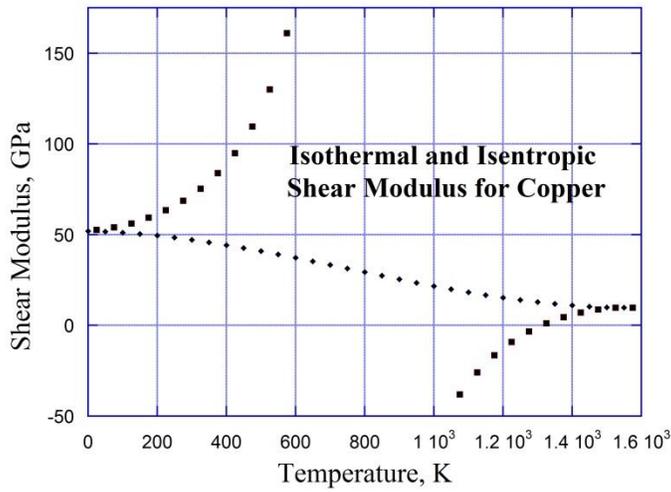

Figure 5. The measured isothermal shear modulus and the isentropic shear modulus constructed from the slope of the constant entropy curve in figure 4 are shown versus $T$. The isothermal data are shown as diamonds and isentropic moduli are plotted as squares; both are shown versus $T$.

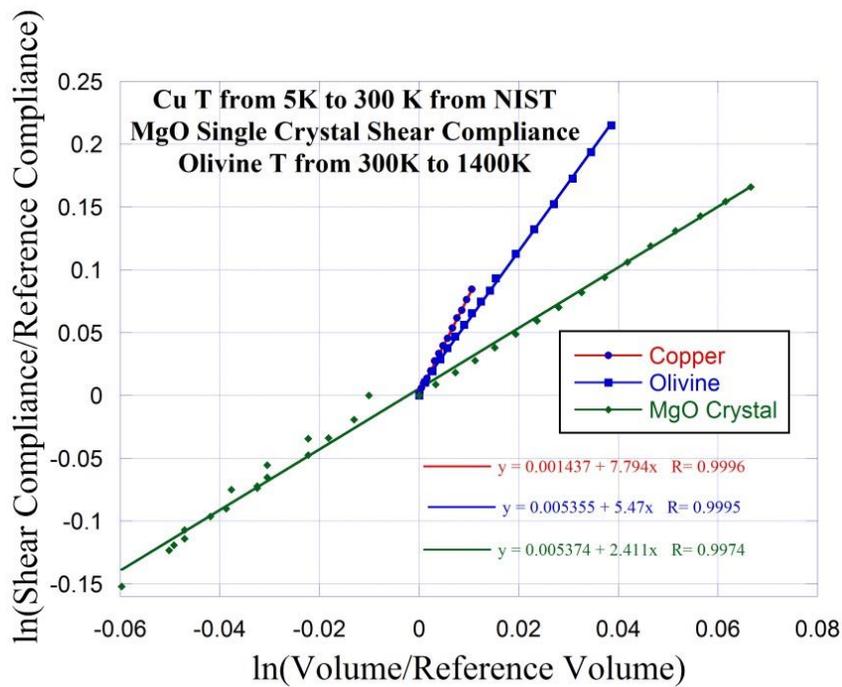

Figure 6. Experimental shear compliances for Copper, Olivine (Fo$_{92}$Fa$_8$) and single crystal MgO plotted as $\ell n$ (shear compliance/reference compliance) versus $\ell n$ (volume/reference volume). The MgO shear compliance is for $S_{44}$ with two points at 0, 0. Data in the first quadrant are from $T$ used as a parameter with each point being a different $T$ value. The data for MgO in the third quadrant are with different pressures at each point. The data are from references [29-33]. Low temperature data for Cu is well below the Debye temperature. For Olivine [33] the shear modulus at 1,000 K was taken as 69.17 GPa being between 70.54 and 67.81 GPa the neighboring points. The table lists this value as 59.17 GPa an assumed typographical error.

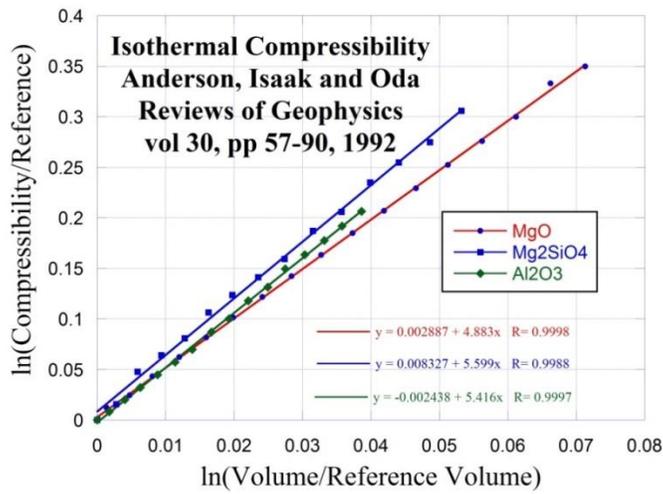

(a)

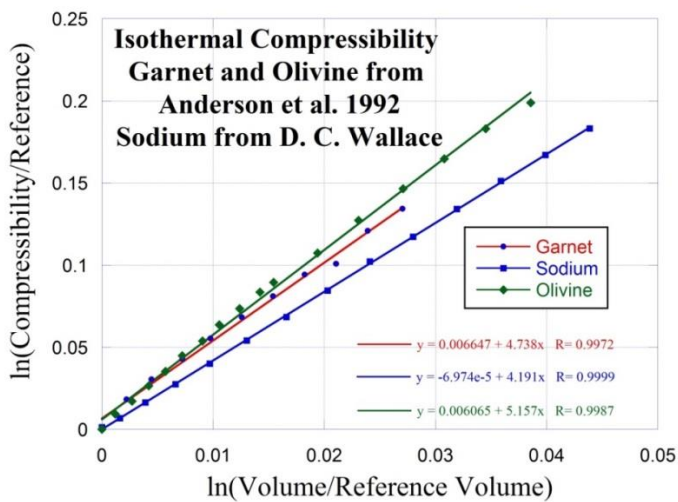

(b)

Figure 7. (a) Experimental data of the isothermal compressibility for polycrystalline magnesia, ringwoodite and alumina [33] plotted as $\ell n$ (compressibility/reference compressibility) versus $\ell n$ (volume/reference volume). (b) Experimental data of the isothermal compressibility for polycrystalline garnet, sodium and olivine [33, 25] is plotted as $\ell n$ (compressibility/reference compressibility) versus $\ell n$ (volume/reference volume). $T$ was used as a parameter for all plots.

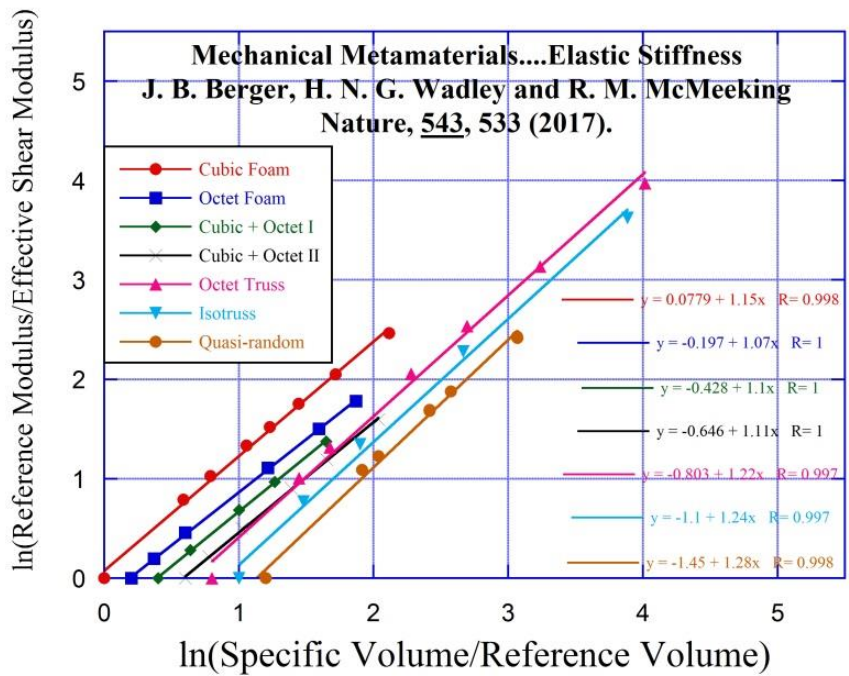

Figure 8. Metamaterial structures analyzed with a finite element program for which representative, isothermal shear moduli and density have been determined [17, 18]. The data are plotted as $\ell n$ (reference/shear modulus) versus $\ell n$ (volume/reference volume). Applied strains were used as a parameter to change the internal shapes in all structures. Each color is a different metastructure curve fit to a power law relation and offset by an increment of 0.2 on the abscissa.

*Table:*

Table 1. A table constructed to describe the thermodynamic system in equation (7). The Table is in Jacobian format and contains the physical properties and Maxwell's equation (9) for the system. All partial derivatives may be found using Jacobian algebra and Table 1 entries. See equation (T-1) and (T2) just below Table 1 for an example.

| function | $\left.\dfrac{\partial}{\partial T}\right|_{\tau}$ | $\left.\dfrac{\partial}{\partial \tau}\right|_{T}$ |
|---|---|---|
| $T$ | 1 | 0 |
| $\tau$ | 0 | 1 |
| $s$ | $\dfrac{C_{\tau}}{T}$ | $\tau \left.\dfrac{\partial \lambda_T}{\partial T}\right|_{\tau}$ |
| $\gamma$ | $\tau \left.\dfrac{\partial \lambda_T}{\partial T}\right|_{\tau}$ | $\lambda_T$ |

$$\left.\frac{\partial \gamma}{\partial \tau}\right|_{s} \equiv \lambda_s = \frac{J(\gamma, s)}{J(\tau, s)} = \frac{\begin{vmatrix} \tau \left.\dfrac{\partial \lambda_T}{\partial T}\right|_{\tau} & \lambda_T \\ C_{\tau}/T & \tau \left.\dfrac{\partial \lambda_T}{\partial T}\right|_{\tau} \end{vmatrix}}{\begin{vmatrix} 0 & 1 \\ C_{\tau}/T & \tau \left.\dfrac{\partial \lambda_T}{\partial T}\right|_{\tau} \end{vmatrix}} = \lambda_T - \frac{T}{C_{\tau}}\left(\tau \left.\frac{\partial \lambda_T}{\partial T}\right|_{\tau}\right)^2. \qquad \text{(T-1)}$$

So
$$\lambda_s = \lambda_T - \frac{T}{C_{\tau}}\left(\tau \left.\frac{\partial \lambda_T}{\partial T}\right|_{\tau}\right)^2. \qquad \text{(T-2)}$$

*References*:

1. Ke, T. S., 1949, "Analysis of the Temperature Coefficient of Shear Modulus of Aluminum," Phys. Rev. (Letters), 76, 579.

2. Lazarus, D., 1949, "The Variation of the Adiabatic Elastic Constants of KCl, NaCl, CuZn, Cu and Al with Pressure to 10,000 Bars," Phys. Rev., 76, pp. 545-553.

3. Grover, R., Getting, I. C. and Kennedy, G. C., 1973, "Simple Compressibility Relation for Solids," Phys. Rev., B7, pp. 567-571.

4. Gibson, L. J., and Ashby, M. F., 1982, "The Mechanics of Three-Dimensional Cellular Materials," Proc. Roy. Soc. Lond., A382, pp.43-59.

5. Ashby, M. F., 1983, "The Mechanical Properties of Cellular Solids," Metall. Trans., A14, pp. 1755-1769.

6. Ashby, M. F., 2006, "The Properties of Foams and Lattices," Phil. Trans. Roy. Soc. London: Math. Phys. Eng. Sci., 364, pp.15-30.

7. Huber, N., 2018 "Connections between Topology and Macroscopic Mechanical Properties of Three-Dimensional Open-Pore Materials," Front. Mater., 5, 069, doi:https://doi.org/10.3389/fmats.2018.00069

8. Anderson, O. L., 1967, "Equation for the Thermal Expansivity in Planetary Interiors," J. Geophysical Res., 72, pp. 3661-3668.

9. Reynard, B., and Price, G. D., 1990, "Thermal Expansion of Mantle Minerals at High Pressure - A Theoretical Study," Geophys. Res. Lett., 17, pp. 689-692.

10. Anderson, O. L., Oda, H., and Isaak, D., 1992, "A Model for the Computation of Thermal Expansivity at High Compression and High Temperatures: MgO as an Example," Geophys. Res. Lett., 19, pp.1987-1990.

11. Barron, T. H. K., 1979, "A Note on the Anderson-Grüneisen Function," J. Phys. C Solid State Physics, 12, pp. L155-L159.

12. Burns, S. J., 2018, "Elastic Shear Modulus Constitutive Law found from Entropy Considerations," J. Appl. Phys., 124, pp., 085904-1-9; https://doi.org/10.1063/1.5041962

13. Burns, S. J., 2018, "77 New Thermodynamic Identities among Crystalline Elastic Material Properties Leading to a Shear Modulus Constitutive Law in Isotropic Solids," J. Appl. Phys., 124, pp. 085114-1-9; https://doi.org/10.1063/1.5041961

14. Birch, F., 1938, "The Effect of Pressure upon the Elastic Parameters of Isotropic Solids, According to Murnaghan's Theory of Finite Strain," J. Appl. Phys., 9, pp. 279-288.

15. Murnaghan, F. D., 1944, "The Compressibility of Media under Extreme Pressures," Proc. Natl. Acad. Sci. USA, 30, pp. 244–247.

16. Qin, Z., Jung, G., Kang, S., and Buehler, M. J., 2017, "The Mechanics and Design of a Lightweight Three-Dimensional Graphene Assembly," Sci. Adv., 3, e1601536.

17. Berger, J. B., Wadley, H. N. G., and McMeeking, R. M., 2017 "Mechanical Metamaterials at the Theoretical Limit of Isotropic Elastic Stiffness," Nature, 543, pp. 533-537.

18. Berger, J. B., Wadley, H. N. G., and McMeeking, R. M., 2018, "Berger et al. reply," Nature 564, pp. E-2-E-4; https://doi.org/10.1038/s41586-018-0725-7 .

19. Milton, G. W., 2018, "Stiff Competition," Nature, 564, E-1; https://doi.org/10.1038/s41586-018-0724-8

20. Hashin, Z., and Shtrikman, S., 1963, "A Variational Approach to the Theory of the Elastic Behavior of Multiphase Materials," J. Mech. Phys. Solids, 11, pp. 127-140.

21. Hall, S. A., Woods, D. A., Ibraim, E., and Viggiani, G., 2009, "Localized Deformation Patterning in 2-D Granular Materials Revealed by Digital Image Correlation," Granular Matter, 12, pp. 1-14.

22. Voigt, W., 1910, Lehrbuch der Kristallphysik, 1st edition (reprinted in 1928 with an additional appendix), Leipzig, Teubner.

23. Nye, J. F., 1964, Physical Properties of Crystals, Oxford Univ. Press, London, UK, pp 131-149.

24. Newnham, R. E., 2004, Properties of Materials, Anisotropy, Symmetry, Structure, Oxford Univ. Press, Oxford, UK.

25. Wallace, D. C., 1972, Thermodynamics of Crystals, John Wiley & Sons, Inc., New York.

26. deWit, R., 2008, "Elastic and Thermal Expansion Averages of a Nontextured Polycrystal," J. Mech. Mater. Struct., 3, pp. 195-212.

27. Kim, K. Y., 1996, "Thermodynamics at Finite Deformation of an Anisotropic Elastic Solid," Phys. Rev., B54, pp. 6245-6254.

28. Barron, T. H. K., 1998, "Generalized Theory of Thermal Expansion in Solids", in CINDAS data series on Materials Properties: V, 1-4, Thermal Expansion in Solids ed., Ho, C. Y., authored by Taylor, R. E. et al., ASM International, Materials Park, OH, USA.

29. Simon, N. J., Drexler, E. S., and Reed, R. P., 1992, Properties of Copper and Copper Alloys at Cryogenic Temperatures, Nat. Inst. Stand. Tech., Monograph 177, pp. 6-11, 7-31 and 7-32.

30. Ledbetter, H. M. and Naimon, E. R., 1974, Elastic Properties of Metals and Alloys. II Copper, J. Phys. Chem., Ref. Data, 3, p. 929.

31. Isaak, D. G., Anderson, O. L., and Goto, T., 1989, "Measured Elastic Moduli of Single-Crystal MgO up to 1800 K," Phys. Chem. Minerals, 16, pp. 704-713.

32. Li, B., Woody, K., and Kung, J., 2006, "Elasticity of MgO to 11 GPa with an independent absolute pressure scale: Implications for pressure calibration," J. Geophys. Res., 111, B11206; doi:10.1029/2005JB004251.

33. Anderson, O. L., Isaak, D., and Oda, H., 1992, "High-temperature elastic constant data on minerals relevant to geophysics," Rev. Geophys., 30, pp. 57-90.

*Appendix A* Derivation of the Elastic Shear Compliance Equation from the Second-Order Differential Equation (16) used in *Universal Law for the Elastic Moduli of Solids and Structures*.

Starting with *S* as the shear compliance in equation (16):

$$\frac{(S')^2}{S\,S''} = \text{constant} = C_1 \qquad (A\text{-}1)$$

This is a second order differential equation that has to be integrated twice to get the general solution, equation (A-1) after rearranging is

$$\frac{S'}{S} = \frac{C_1 S''}{S'}. \qquad (A\text{-}2)$$

Both sides can be integrated to yield:

$$\frac{1}{C_1} \ln(S) = \ln(S') + \ln(C_2). \qquad (A\text{-}3)$$

$C_2$ is the first constant of integration. Equation (A-3) after rearranging terms is:

$$\ln\left(\frac{S^{1/C_1}}{S'}\right) = \ln(C_2). \qquad (A\text{-}4)$$

so

$$\frac{S^{1/C_1}}{S'} = C_2. \qquad (A\text{-}5)$$

After rearranging equation (A-5) we have:

$$C_2 \frac{dS}{S^{1/C_1}} = dv. \qquad (A\text{-}6)$$

Equation (A-6) can be integrated yielding:

$$\frac{C_2}{(1 - 1/C_1)} S^{(1 - 1/C_1)} = v + C_3. \qquad (A\text{-}7)$$

$C_3$ is the second constant of integration. The reference state of $S_0$ and $v_0$ determines $C_3$ using equation (A-7).

$$C_3 = \frac{C_2}{(1 - 1/C_1)} S_0^{(1 - 1/C_1)} - v_0. \qquad (A\text{-}8)$$

Substitution of $C_3$ into equation (A-7) yields an expression for the shear compliance S as it relates to the volume, $v$. After rearrangement, we have from equation (A-7) with equation (A-8) eliminating $C_3$:

$$\frac{C_2}{(1-1/C_1)} S_0^{(1-1/C_1)} \left\{ 1 - \left(\frac{S}{S_0}\right)^{(1-1/C_1)} \right\} = v_0(1 - \frac{v}{v_0}). \qquad (A-9)$$

Expression (A-9) is simplified with the constants $m$ and $a$.

$$\left(1 - \left(\frac{S}{S_0}\right)^{1/m}\right) = a(1 - \frac{v}{v_0}), \qquad (A-10)$$

with

$$m = \frac{C_1}{(C_1 - 1)} \text{ and } a = \frac{v_0}{C_2} \frac{1 - 1/C_1}{S_0^{(1-1/C_1)}}. \qquad (A-11)$$

Chose $C_2$ so $a = 1$. Expression (A-10) is thus a power law between $S$ and $v$ as given below.

The value of $m$ can be determined from experimental data to relate $S$ to $v$ with equation (A-12):

$$\ell n\left(\frac{S}{S_0}\right) = m \, \ell n\left(\frac{v}{v_0}\right). \qquad (A-12)$$

Reference [3] uses an exponential function to relate the bulk modulus to $v$. This is an engineering approximation to the volumetric strain as seen in equation (A-13).

$$\ell n(\frac{v}{v_0}) = \ell n(\frac{v_0 + \Delta}{v_0}) = \ell n(1 + \frac{\Delta v}{v_0}) \approx \frac{\Delta v}{v_0} - \frac{1}{2}\left(\frac{\Delta v}{v_0}\right)^2 + \ldots\ldots (A-13)$$

Keeping only the first term on the right of expansion in (A-13) and using (A-12) it is found that

$$\ell n(\frac{S}{S_0}) = m \frac{\Delta v}{v_0} \qquad (A-14)$$

$$S = S_0 e^{m\frac{\Delta v}{v_0}} \qquad (A-15)$$

The shear compliance from equation (A-15) is thus in the same form as the reciprocal bulk modulus in [3] provided $\frac{\Delta v}{v_0}$ is not too large.